%% file: arXclmc.tex
\documentclass[sts]{imsart}

\usepackage[utf8]{inputenc}
\usepackage{amsmath,amssymb,amsthm}
\usepackage{graphicx}
\usepackage{caption}
\usepackage{color}
\usepackage{bm}
\usepackage{float}
\usepackage{tipa}
\usepackage{comment}
\usepackage{natbib}
\usepackage{nicefrac}
\definecolor{background-color}{gray}{0.98}

\begin{document}

\begin{frontmatter}
\title{Accelerating MCMC Algorithms}

\begin{aug}
\author{\snm{Christian P Robert}}
\affiliation{Universit\'e Paris Dauphine, PSL Research
University, and Department of Statistics, University of Warwick}
\author{\snm{V\'ictor Elvira}}
\affiliation{IMT Lille Douai \& CRIStAL laboratory} 
\author{\snm{Nick Tawn}}
\affiliation{Department of Statistics, University of Warwick}
\author{\snm{Changye Wu}}
\affiliation{Universit\'e Paris Dauphine, PSL Research University}
\end{aug}

\begin{abstract}
Markov chain Monte Carlo algorithms are used to simulate from complex statistical
distributions by way of a local exploration of these distributions. This local
feature avoids heavy requests on understanding the nature of the target, but it
also potentially induces a lengthy exploration of this target, with a
requirement on the number of simulations that grows with the dimension of the
problem and with the complexity of the data behind it. Several techniques are
available towards accelerating the convergence of these Monte Carlo algorithms,
either at the exploration level (as in tempering, Hamiltonian Monte Carlo and
partly deterministic methods) or at the exploitation level (with
Rao-Blackwellisation and scalable methods).
\end{abstract}
\end{frontmatter}


\section{Introduction}

Markov chain Monte Carlo (MCMC) algorithms have been used for nearly 60 years,
becoming a reference method for analysing Bayesian complex models in the early
1990's \citep{gelfand:smith:1990}. The strength of this method is that it
guarantees convergence to the quantity (or quantities) of interest with minimal
requirements on the targeted distribution (also called {\em target}) behind such quantities. In that sense, MCMC
algorithms are robust or universal, as opposed to the most standard Monte Carlo methods
\citep[see, e.g.,][]{rubinstein81, robert:casella:2004} that require direct
simulations from the target distribution. This robustness may however induce a
slow convergence behaviour in that the exploration of the relevant
space---meaning the part of the space supporting the distribution that has a
significant probability mass under that distribution---may take a long while, as the
simulation usually proceeds by local jumps in the vicinity of the current
position. In other words, MCMC--especially in its off-the-shelf versions like
Gibbs sampling and Metropolis--Hastings algorithms---is very often myopic in
that it provides a
good illumination of a local area, while remaining unaware of the global
support of the distribution. As with most other
simulation methods, there always exist ways of creating highly convergent MCMC
algorithms by taking further advantage of the structure of the target
distribution. Here, we mostly limit ourselves to the realistic situation where
the target density is only known as the output of a computer code or to a setting
similarly limited in its information content.

The approaches to the acceleration of MCMC algorithms can be divided in several categories, from those
which improve our knowledge about the target distribution, to those that modify
the proposal in the algorithm, including those that exploit better the outcome of the
original MCMC algorithm. The following sections provide more details about
these directions and the solutions proposed in the literature.

\section{What is MCMC and why does it need accelerating?}\label{sec:why}

MCMC methods have a history \citep[see, e.g.][]{cappe:robert:2000b} that
starts at approximately the same time as the Monte Carlo methods, in
conjunction with the conception of the first computers. They have been devised
to handle the simulation of complex target distributions, when complexity stems
from the shape of the target density, the size of the associated data, the
dimension of the object to be simulated, or from time requirements. For instance,
the target density $\pi(\theta)$ may happen to be expressed in terms of
multiple integrals that cannot be solved analytically,
$$
\pi(\theta) = \int \omega(\theta,\xi)\text{d}\xi\,,
$$ 
which requires the simulation of the entire vector $(\theta,\xi)$. In cases
when $\xi$ is of the same dimension as the data, as for instance in latent
variable models, this significant increase in the dimension of the object to be
simulated creates computational difficulties for standard Monte Carlo methods,
from managing the new target $\omega(\theta,\xi)$, to devising a new and
efficient simulation algorithm. A Markov chain Monte Carlo (MCMC) algorithm
allows for an alternative resolution of this computational challenge by
simulating a Markov chain that explores the space of interest (and possibly
supplementary spaces of auxiliary variables) without requiring a deep
preliminary knowledge on the density $\pi$, besides the ability to compute
$\pi(\theta_0)$ for a given parameter value $\theta_0$ (if up to a normalising
constant) and possibly the gradient $\nabla \log \pi(\theta_0)$. The validation
of the method \citep[e.g.,][]{robert:casella:2004} is that the Markov chain is
{\em ergodic} \citep[e.g.,][]{meyn:tweedie:1993}, namely that it converges in
distribution to the distribution with density $\pi$, no matter where the Markov
chain is started at time $t=0$. 

The Metropolis--Hastings algorithm is a generic illustration of this principle. The
basic algorithm is constructed by choosing a {\em proposal}, that is, a conditional
density $K(\theta'|\theta)$ (also known as a {\em Markov kernel}), the Markov
chain $\{\theta_t \}_{t=1}^{\infty}$ being then derived by successive simulations of the
transition
$$\theta_{t+1} = \begin{cases}
\theta^\prime \sim K(\theta^\prime |\theta_t) &\text{with probability }
\left\{\dfrac{\pi(\theta^\prime)}{\pi(\theta_t)}\times\dfrac{K(\theta_t|\theta^\prime)}
{K(\theta^\prime|\theta_t)}\right\}\wedge 1,\\
\theta_t &\text{otherwise}.\\
\end{cases}$$
This acceptance-rejection feature of the algorithm makes it appropriate for
targeting $\pi$ as its stationary distribution if the resulting Markov chain
$\{\theta_t\}_{t=1}^{\infty}$ is irreducible, i.e., has a positive probability of visiting any
region of the support of $\pi$ in a finite number of iterations. (Stationarity
can easily be shown, e.g., by using the so-called {\em detailed balance
property} that makes the chain time-reversible, \citealp[see, e.g.,][]{robert:casella:2004}.) 

Considering the initial goal of simulating samples from the target distribution $\pi$,
the performances of MCMC methods like the Metropolis--Hastings algorithm above
often vary quite a lot, depending primarily on the correspondance between the proposal $K$ and the
target $\pi$. For instance, if $K(\theta|\theta_t)=\pi(\theta)$, the
Metropolis--Hastings algorithm reduces to i.i.d.~sampling from the target,
which is of course a formal option when i.i.d.~sampling from $\pi$ proves impossible to implement.
Although there exist rare instances when the Markov chain $\{\theta_t\}_{t=1}^\infty$ leads
to negative correlations between the successive terms of the chain, making it
{\em more efficient} than regular i.i.d.~sampling \citep{liu:won:kon95}, the
most common occurrence is one of positive correlation between the simulated
values (sometimes uniformly, see \citealp{liu:won:kon94}). This feature implies a
reduced efficiency of the algorithm and hence requires a larger number of
simulations to achieve the same precision as an approximation based
on i.i.d.~simulations (without accounting for differences in computing time). More
generally, a MCMC algorithm may require a large number of iterations to escape
the attraction of its starting point $\theta_0$ and to reach stationarity, to the extent that
some versions of such algorithms fail to converge in the time available (i.e.,
in practice if not in theory). 

It thus makes sense to seek ways of accelerating (a) the convergence of a given
MCMC algorithm to its stationary distribution, (b) the convergence of a given
MCMC estimate to its expectation, and/or (c) the exploration of a given MCMC
algorithm of the support of the target distribution. Those goals are related
but still distinct. For instance, a chain initialised by simulating from the target
distribution may still fail to explore the whole support in an acceptable
number of iterations. While there is not an optimal and universal solution to
this issue, we will discuss below approaches that are as generic as possible,
as opposed to artificial ones taking advantage of the mathematical structure of
a specific target distribution. Ideally, we aim at covering realistic
situations when the target density is only known [up to a constant or an
additional completion step] as the output of an existing computer code.
Pragmatically, we also cover here solutions that require more efforts and
calibration steps when they apply to a wide enough class of problems.

\section{Accelerating MCMC by exploiting the geometry of the target}\label{sec:geo}

While there is no end in trying to construct more efficient and faster MCMC
algorithms, and while this (endless) goal needs to account for the cost of devising such
alternatives under limited resources budgets, there exist several generic
solutions such that a given target can first be explored in terms of the
geometry (or topology) of the density before constructing the algorithm. Although
this type of methods somehow takes us away from our original purpose which was
to improve upon an existing algorithm, they still make sense within this survey in
that they allow for almost automated implementations. 

\subsection{Hamiltonian Monte Carlo}

From the point of view of this review, Hamiltonian (or hybrid) Monte Carlo
(HMC) is an auxiliary variable technique that takes advantage of a continuous
time Markov process to sample from the target $\pi$. This approach comes from
physics \citep{duane:etal:1987} and was popularised in statistics by
\cite{neal:1996,neal:2011} and \cite{mackay:2002}. Given a target
$\pi(\theta)$, where $\theta\in\mathbb{R}^d$, an artificial auxiliary variable
$\vartheta\in\mathbb{R}^d$ is introduced along with a density $\varpi(\vartheta|\theta)$
so that the joint distribution of $(\theta,\vartheta)$ enjoys $\pi(\theta)$ as
its marginal. While there is complete freedom in this representation, the HMC
literature often calls $\vartheta$ the {\em momentum} of a particle located at $\theta$ by analogy
with physics. Based on the representation of the joint distribution
$$
\omega(\theta,\vartheta)
=\pi(\theta)\varpi(\vartheta|\theta) \propto \exp\{ -H(\theta,\vartheta) \}\,,
$$
where $H(\cdot)$ is called the {\em Hamiltonian}, Hamiltonian Monte Carlo (HMC)
is associated with the continuous time process
$(\theta_t,\vartheta_t)$ generated by the so-called {\em Hamiltonian equations}
$$
\dfrac{\text{d}\theta_t}{\text{d}t}=\dfrac{\partial H}{\partial \vartheta}(\theta_t,\vartheta_t)\qquad
\dfrac{\text{d}\vartheta_t}{\text{d}t}=-\dfrac{\partial H}{\partial \theta}(\theta_t,\vartheta_t)\,,
$$
which keep the Hamiltonian target stable over time, as
$$
\dfrac{\text{d}H(\theta_t,\vartheta_t)}{\text{d}t}=\dfrac{\partial H}{\partial \vartheta}(\theta_t,\vartheta_t)\,\dfrac{\text{d}\vartheta_t}{\text{d}t}+\dfrac{\partial H}{\partial \theta}(\theta_t,\vartheta_t)\,\dfrac{\text{d}\theta_t}{\text{d}t}=0\,.
$$
Obviously, the above continuous time Markov process is deterministic and only
explores a given level set, $$\{(\theta, \vartheta) : H(\theta, \vartheta) =
H(\theta_0, \vartheta_0)\}\,,$$ instead of the whole augmented state space
$\mathbb{R}^{2d}$, which induces an issue with irreducibility. An acceptable solution 
to this problem is to refresh the momentum, $\vartheta_{t} \sim \varpi(\vartheta|\theta_{t-})$, at
random times $\{\tau_n\}_{n=1}^{\infty}$, where $\theta_{t-}$ denotes the 
location of $\theta$ immediately prior to time $t$, and the random 
durations $\{\tau_n - \tau_{n-1}\}_{n=2}^{\infty}$ follow an exponential
distribution. By construction, continuous-time Hamiltonian Markov chain can be
regarded as a specific  piecewise deterministic Markov process using
Hamiltonian dynamics \citep{davis1984piecewise, davis1993markov,
bou2017randomized}  and our target, $\pi$, is the marginal of its associated
invariant distribution.

Before moving to the practical
implementation of the concept, let us point out that the free cog in the
machinery is the conditional density $\varpi(\vartheta|\theta)$, which is
usually chosen as a Gaussian density with either a constant covariance matrix $M$
corresponding to the target covariance or as a local curvature depending on
$\theta$ in Riemannian Hamiltonian Monte Carlo \citep{girolami:2011}.
\cite{betancourt:2017} argues in favour of these two cases against non-Gaussian
alternatives and \cite{livingstone2017kinetic} analyse how different choices
of kinetic energy in Hamiltonian Monte Carlo affect algorithm performances. For
a fixed covariance matrix, the Hamiltonian equations become
$$
\dfrac{\text{d}\theta_t}{\text{d}t}=
M^{-1}\vartheta_t\qquad \dfrac{\text{d}\vartheta_t}{\text{d}t}=\nabla \mathcal{L}(\theta_t)\,,
$$
which is the score function. The velocity (or momentum) of the process is thus
driven by this score function, gradient of the log-target.

The above description remains quite conceptual in that there is no generic
methodology for producing this continuous time process, since the Hamiltonian
equations cannot be solved exactly in most cases. Furthermore, standard
numerical solvers like Euler's method create an instable approximation that
induces a bias as the process drifts away from its true trajectory. There
exists however a discretisation simulation technique that produces a Markov
chain and is well-suited to the Hamiltonian equations in that it preserves the
stationary distribution \citep{betancourt:2017}. It is called the {\em
symplectic integrator}, and one version in the independent case with constant
covariance consists in the following (so-called {\em leapfrog}) steps
\begin{align*}
\vartheta_{t+\epsilon/2} &= \vartheta_t+\epsilon \nabla \mathcal{L}(\theta_t)/2,\\
\theta_{t+\epsilon} &= \theta_t+\epsilon M^{-1} \vartheta_{t+\epsilon/2},\\
\vartheta_{t+\epsilon} &= \vartheta_{t+\epsilon/2}+\epsilon \nabla \mathcal{L}(\theta_{t+\epsilon})/2,
\end{align*}
where $\epsilon$ is the time-discretisation step. Using a proposal on
$\vartheta_0$ drawn from the Gaussian auxiliary target and deciding on the
acceptance of the value of $(\theta_{T\epsilon},\vartheta_{T\epsilon})$ by a
Metropolis--Hastings step can limit the danger of missing the
target. Note that the first two leapfrog steps induce a Langevin move on
$\theta_t$:
$$
\theta_{t+\epsilon} = \theta_t+\epsilon^2 M^{-1} \nabla \mathcal{L}(\theta_t)/2
+\epsilon M^{-1} \vartheta_{t}\,,
$$
thus connecting with the MALA algorithm discussed below (see \citealp{durmus:moulines:2017} 
for a theoretical discussion of the optimal choice of $\epsilon$). Note that the leapfrog integrator
is quite an appealing middleground between accuracy (as it is second-order
accurate) and computational efficiency.

In practice, it is important to note that discretising the Hamiltonian dynamics
introduces two free parameters, the step size $\epsilon$ and the trajectory
length $T\epsilon$, both to be calibrated. As an empirically successful and popular
variant of HMC, the ``no-U-turn sampler'' (NUTS) of \cite{hoffman2014no} adapts
the value of $\epsilon$ based on primal-dual averaging. It also eliminates the
need to choose the trajectory length $T$ via a recursive algorithm that builds
a set of candidate proposals for a number of forward and backward leapfrog
steps and stops automatically when the simulated path steps back. 

A further acceleration step in this area is proposed by \cite{rasmussen:2003}
\citep[see also][]{fielding:nott:liong:2011}, namely the replacement of the
exact target density $\pi(\cdot)$ by an approximation $\hat\pi(\cdot)$ that is
much faster to compute in the many iterations of the HMC algorithm. A generic
way of constructing this approximation is to rely on Gaussian processes, when
interpreted as prior distributions on the target density $\pi(\cdot)$, which is
only observed at some values of $\theta$, $\pi(\theta_1),\ldots,\pi(\theta_n)$
\citep{rasmussen:wiliams:2006}. This solution is speeding up the algorithm,
possibly by orders of magnitude, but it introduces a further approximation into
the Monte Carlo approach, even when the true target is used at the end of the
leapfrog discretisation, as in \cite{fielding:nott:liong:2011}.

\begin{figure}[h]
\centering
\includegraphics[width=1\textwidth]{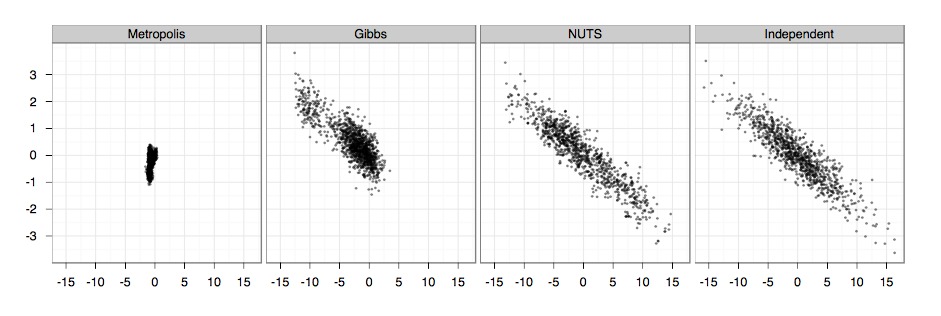}
\caption{
Comparisons between random-walk Metropolis-Hastings, Gibbs sampling, and NUTS 
algorithm of samples corresponding to a highly correlated 250-dimensional
multivariate Gaussian target. Similar computation budgets are used for all methods to produce the
1,000 samples on display ({\em Source:} \cite{hoffman2014no}, with permission).
}
\label{fig:nuts}
\end{figure}

Stan (named after Stanislas Ullam, see \citealp{carpenter:etal:2017}) is a computer language for 
Bayesian inference that, among other approximate techniques, implements the NUTS
algorithm to remove hand-tuning.  More precisely, Stan is a probabilistic
programming language in that the input is at the level of a statistical model,
along with data, rather than the specifics of an MCMC algorithm. The algorithmic
part is somehow automated, meaning that when models can be conveniently
defined through this language, it offers an alternative to the sampler that
produced the original chain. As an illustration of the acceleration brought by
HMC, Figure~\ref{fig:nuts}, reproduced from \cite{hoffman2014no}, shows the
performance of NUTS, compared with both random-walk MH and Gibbs samplers.

\section{Accelerating MCMC by breaking the problem into pieces}\label{sec:scal}

The explosion in the collection and analysis of ``big'' datasets in recent years has
brought new challenges to the MCMC algorithms that are used for Bayesian inference. When examining
whether or not a new proposed sample is accepted at the accept-reject step, an MCMC algorithm such as the
Metropolis-Hastings version needs to sweep over the whole data set, at each
and every iteration, for the evaluation of the likelihood function. MCMC algorithms are then difficult to scale up, which strongly hinders their
application in big data settings. In some cases, the datasets may be too large
to fit on a single machine. It may also be that confidentiality measures impose
different databases to stand on separate networks, with the possible added
burden of encrypted data \citep{aslet:esperanza:holmes:2015}. Communication
between the separate machines may prove impossible on an MCMC scale that
involves thousands or hundreds of thousands iterations.

\subsection{Scalable MCMC methods}

In the recent years, efforts have been made to design {\em scalable} algorithms, namely, solutions that manage to handle large scale targets by breaking the problem into manageable or scalable pieces.
Roughly speaking, these methods can be classified into two categories
\citep{bardenet2015markov}: divide-and-conquer approaches and sub-sampling approaches.

Divide-and-conquer approaches partition the whole data set, denoted
$\mathcal{X}$, into batches, $\{\mathcal{X}_1, \cdots, \mathcal{X}_k\}$, and run
separate MCMC algorithms on each data batch, independently, as if they were independent
Bayesian inference problems.\footnote{In order to keep the notations
consistent, we still denote the target density by $\pi$, with the prior density
denoted as $\pi_0$ and the sampling distribution of one observation $x$ as
$p(x|\theta)$. The dependence on the sample $\mathcal{X}$ is not reported
unless necessary.} These methods then combine the simulated parameter outcomes together
to approximate the original posterior distribution. Depending on the
treatments of the batches selected in the MCMC stages, these approaches can be
further subdivided into two finer groups: sub-posterior methods and boosted
sub-posterior methods. Sub-posterior methods are motivated by the
independent product equation:
\begin{equation}
\pi(\theta) \propto
\prod_{i=1}^k\left(\pi_0(\theta)^{1/k}\prod_{\ell\in\mathcal{X}_i}p(x_{\ell}|\theta)\right)
=\prod_{i=1}^k\pi_i(\theta)\,,
\end{equation}
and they target the densities $\pi_i(\theta)$ (up to a constant) in their
respective MCMC steps. They thus bypass communication costs \citep{scott2016bayes}, by running MCMC
samplers independently on each batch, and they most often increase MCMC mixing
rates (in effective samples sizes produced by second), given that the
sub-posterior distributions $\pi_i(\theta)$ are based on smaller datasets. For
instance, \cite{scott2016bayes} combine the samples from the sub-posteriors,
$\pi_i(\theta)$, by a Gaussian reweighting.
\cite{neiswanger2013asymptotically} estimate the sub-posteriors $\pi_i(\theta)$
by non-parametric and semi-parametric methods, and they run additional MCMC
samplers on the product of these estimators towards approximating the true
posterior $\pi(\theta)$.  \cite{wang2013parallelizing} refine this product
estimator with an additional Weierstrass sampler, while
\cite{wang2015parallelizing} estimate the posterior by partitioning the space
of samples with step functions. \cite{vehtari:etal:2014} devised an expectation propagation scheme
to improve the postprocessing of the parallel samplers. 

As an alternative to sampling from the sub-posteriors, boosted sub-posterior
methods target instead the components
\begin{equation}
\tilde{\pi}_i(\theta) \propto \pi_0(\theta)\left(\prod_{\ell\in\mathcal{X}_i}p(x_{\ell}|\theta)\right)^k
\end{equation}
in separate MCMC runs. Since they formaly amount to repeating each batch $k$
times towards producing pseudo data sets with the same size as the true
one, the resulting boosted sub-posteriors, $\tilde{\pi}_1(\theta), \cdots,
\tilde{\pi}_k(\theta)$, have the same scale in variance of each component of
the parameters, $\theta$, as the true posterior, and can thus be treated as a
group of estimators of the true posterior.  In the subsequent combining stage,
these sub-posteriors are merged together to construct a better approximation of the
target distribution. For instance, \cite{minsker2014scalable} approximate the
posterior with the geometric median of the boosted sub-posteriors, embedding
them into associated reproducing kernel Hilbert spaces (rkhs), while
\cite{srivastava2015wasp} achieve this goal using the barycentres of
$\tilde{\pi}_1,\cdots, \tilde{\pi}_k$, these barycentres being computed with
respect to a Wasserstein distance.

In a perspective different from the above parallel scheme of divide-and-conquer approaches,
sub-sampling approaches aim at reducing the number of individual datapoint
likelihood evaluations operated at each iteration towards accelerating MCMC algorithms.
From a general perspective, these approaches can be further classified into two finer classes:
exact subsampling methods and approximate subsampling methods, depending on their
resulting outputs. Exact subsampling approaches typically require subsets of
data of random size at each iteration. One solution to this effect is taking advantage of
pseudo-marginal MCMC via constructing unbiased estimators of 
the target density evaluated on subsets of the data
\citep{andrieu2009pseudo}. \cite{quiroz2016exact} follow this direction by
combining the powerful debiasing technique of \cite{rhee2015unbiased} and the correlated
pseudo-marginal MCMC approach of \cite{deligiannidis2015correlated}. Another direction is to
use piecewise deterministic Markov processes (PDMP)
\citep{davis1984piecewise,davis1993markov}, which enjoy the target
distribution as the marginal of their invariant distribution. This PDMP version requires
unbiased estimators of the gradients of the log-likelihood function,
instead of the likelihood itself. By using a tight enough bound on the event
rate function of the associated Poisson processes, PDMP can produce
super-efficient scalable MCMC algorithms. The bouncy particle sampler
\citep{bouchard2017bouncy} and the zig-zag sampler \citep{bierkens2016zig} are
two competing PDMP algorithms, while \cite{bierkens2017piecewise} unify and
extend these two methods. Besides, one should note that PDMP produces a
non-reversible Markov chain, which means that the algorithm should be more
efficient in terms of mixing rate and asymptotic variance, when compared with
reversible MCMC algorithms, such as MH, HMC and MALA, as observed in some
theoretical and experimental works \citep{hwang1993accelerating,
sun2010improving,chen2013accelerating,bierkens2016non}.

Approximate subsampling approaches aim at constructing an approximation of the
target distribution. Beside the aforementioned attempts of \cite{rasmussen:2003}
and \cite{fielding:nott:liong:2011}, one direction is to approximate the acceptance probability
with high accuracy by using subsets of the data
\citep{bardenet2014towards,bardenet2015markov}. {Another solution is based on a
direct modification of exact methods. The seminal work
of \cite{welling2011bayesian}, SGLD, is to exploit the Langevin diffusion 
\begin{equation}
\text{d}\mathbf{\theta}_t = \frac{1}{2}\pmb{\Lambda}\nabla\log\pi(\mathbf{\theta}_t)\text{d}t + \pmb{\Lambda}^{1/2}\text{d}\mathbf{B}_t,\quad \mathbf{\theta}_0 \in\mathbb{R}^d, t\in[0, \infty)
\end{equation} 
where $\pmb{\Lambda}$ is a user-specified matrix, $\pi$ is the target
distribution and $\mathbf{B}_t$ is a d-dimensional Brownian process. By virtue of the
Euler-Maruyama discretisation and using unbiased estimators of the gradient of
the log-target density, SGLD and its variants
\citep{ding2014bayesian,chen2014stochastic} often produce fast and accurate results
in practice when compared with MCMC algorithms using MH steps.} 

\begin{figure}[h]
\centering
\includegraphics[width=1\textwidth]{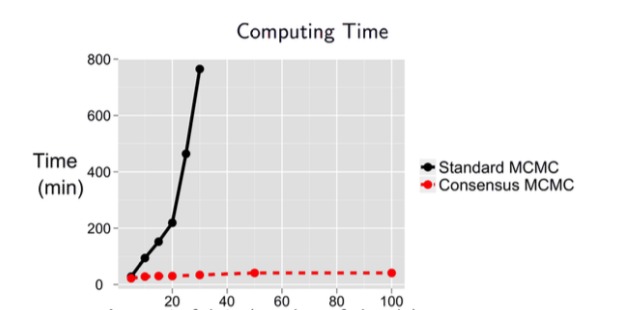}
\caption{Elapsed time when drawing 10,000 MCMC samples with different amounts of data under
     the single machine and consensus Monte Carlo algorithms for a hierarchical Poisson regression.
     The horizontal axis represents the amounts of data. The single machine algorithm stops after
     30 because of the explosion in computation budget.
      ({\em Source:} \cite{scott2016bayes}, with permission.)}
\label{fig:cmc}
\end{figure}

\begin{figure}[h]
\centering
\includegraphics[width=1\textwidth]{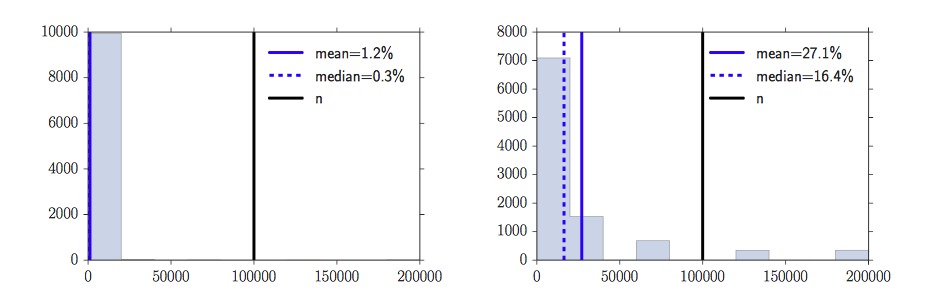}
\caption{Percentage of numbers of data points used in each iteration of the confidence sampler with a
     single 2nd order Taylor approximation at $\theta_{\text{MAP}}$. The plots describe 10,000 iterations
     of the confidence sampler for the posterior distribution of the mean and variance of a uni-dimensional
     Normal distribution with a flat prior: {\em (left)} 10,000 observations are generated from 
     $\mathcal{N}(0,1)$, {\em (right)} 10,000
     observations are generated from $\mathcal{LN}(0,1)$ ({\em Source:}
     \cite{bardenet2015markov}, with permission).}
\label{fig:subsampling}
\end{figure}

Figure~\ref{fig:cmc} shows the time requirements of a consensus Monte Carlo
algorithm \citep{scott2016bayes} compared with a Metropolis--Hastings algorithm using the whole
dataset, while Figure~\ref{fig:subsampling} displays the saving in
likelihood evaluations in confidence sampler of \cite{bardenet2015markov}.

\subsection{Parallelisation and distributed schemes}

Modern computational architectures are built with several computing units that
allow for parallel processing, either fully independent or with certain
communication. Although the Markovian nature of MCMC is inherently sequential
and somewhat alien to the notion of parallelising, several partial solutions
have been proposed in the literature for exploiting these parallel
architectures.  The simplest approach consists in running several MCMC chains
in parallel, blind to all others, until the allotted computing time is
exhausted. Finally, the resulting estimators of all chains are averaged.
However, this naive implementation may suffer from the fact that some of those
chains have not reached their stationary regime by the end of the computation
time, which then induces a bias in the resulting estimate.  Ensuring that
stationarity has been achieved is a difficult (if at all possible) task,
although several approaches can be found in the literature
\citep{mykland1995regeneration,guihenneuc1998discretization,jacob:oleary:atchade:2017}.  At the opposite
extreme, complex targets may be represented as products that involve many terms
that must be evaluated, each of which can be attributed to a different thread
before being multiplied all together.  This strategy requires communication
among processors at each MCMC step.  A middle-ground version
\citep{jacob2011using} consists in running several Markov chains in parallel
with periodic choices of the reference chain, all simulations being recycled
through a Rao-Blackwell scheme. (See also \citealp{calderhead2014general} for a similar scheme.)  
The family of interacting \emph{orthogonal}
MCMC methods (O-MCMC) is proposed in \cite{martino2016orthogonal} with the aim
of fostering better exploration of the state space, specially in
high-dimensional and multimodal targets.  Multiple MCMC chains are run in
parallel exploring the space with random-walk proposals.  The parallel chains
periodically share information, also through joint MCMC steps, thus allowing an
efficient combination of global (coordinated) exploration and local
approximation.  O-MCMC methods also allow for a parallel implementation of the
Multiple Try Metropolis (MTM).  In \cite{calderhead2014general}, a
generalisation of the Metropolis-Hastings algorithm allows for a
straightforward parallelisation.  Each proposed point can be evaluated in a
different processor at every MCMC iteration.  Finally, note that the 
section on scalable MCMC also contains parallelisable approaches, such as the
prefetching method of \cite{angelino:etal:2014} (see also
\citealp{banterle2015accelerating} for a related approach, primarily based on
an approximation of the target).  A most recent endeavour called asynchronous
MCMC \citep{terenin:simpson:draper:2015} aims at higher gains in
parallelisation by reducing the amount of exchange between the parallel
threads, but the notion still remains confidential at this stage.

\section{Accelerating MCMC by improving the proposal}\label{sec:newprop}

In the same spirit as the previous section, this section is stretching the
purpose of this paper by considering possible modifications of the MCMC
algorithm itself, rather than merely exploiting the output of a given MCMC
algorithm. For instance, devising an HMC algorithm is an answer to this
question even though the ``improvement'' is not garanteed. Nonetheless, our
argument here is that, once provided with this output, it is possible to derive
new proposals in a semi-autonomous manner.

\subsection{Simulated tempering}

The target distribution, $\pi(\theta)$ on $d$-dimensional state space $\Theta$,
can exhibit multi-modality with the probability mass being located in different
regions in the state space. The majority of MCMC algorithms use a localised
proposal mechanism which is tuned towards local approximate optimality see,
e.g.,\ \cite{roberts1997weak} and \cite{roberts2001optimal}. By construction,
these localised proposals result in the Markov chain becoming ``trapped'' in a
subset of the state space meaning that in finite run-time the chain can
entirely fail to explore other modes in the state space, leading to biased
samples. Strategies to accelerate MCMC often use local gradient information and
this draws the chain back towards the centre of the mode, which is the opposite
of what is required in a multi-modal setting.

There is an array of methodology available to overcome issues of multi-modality
in MCMC,  the majority of which use state space augmentation. Auxiliary
distributions that allow a Markov chain to explore the entirety of the state
space are targeted and their mixing information is then passed on to aid mixing
in the true target. While the sub-posteriors of the previous section can be
seen as special cases of the following, the most successful and convenient implementation of
these methods is to use {\em power-tempered target distributions}. The target
distribution at inverse temperature level, $\beta$, for $\beta \in (0,1]$ is defined as
\[\pi_\beta(\theta)=\mathfrak{K}(\beta)\left[\pi(\theta)\right]^\beta~~~\mbox{where}~~~\mathfrak{K}(\beta)=\left[\int
\left[\pi(\theta)\right]^\beta d\theta\right]^{-1}.\]
Therefore, $\pi_1(\theta)=\pi(\theta)$.
Temperatures $\beta<1$ flatten out the target distribution allowing the chain to
explore the entire state space provided the $\beta$ value is sufficiently
small. The simulated tempering (ST) and parallel tempering (PT) algorithms
\citep{geyer1991markov, marinari1992simulated} typically use the power-tempered
targets to overcome the issue of multi-modality. The ST approach runs a single
Markov chain on the augmented state space $\{ B, \Theta\}$, where $B=
\{\beta_0,\beta_1,\ldots,\beta_n \}$ is a discrete collection of $n$ inverse
temperature levels with $1=\beta_0>\beta_1>\ldots>\beta_n>0$. The algorithm
uses a Metropolis-within-Gibbs strategy by cycling between updates in the
$\Theta$ and $B$ components of the space. For instance, a proposed temperature swap
move $\beta_i \rightarrow \beta_j$ is accepted with probability  
$$ 
\min \left\{ 1, \frac{\pi_{\beta_j}(\theta)}{\pi_{\beta_i}(\theta)}  \right\} 
$$
in order to preserve detailed balance. Note that this acceptance ratio depends on
the normalisation constants $\mathfrak{K}(\beta)$ which are typically unknown, although
they can sometimes be estimated, as in, e.g.,\ \cite{Wang2001a} and \cite{Atchade2004}. In
case estimation of the marginal normalisation constants is
impractical then the PT algorithm is employed. This approach simultaneously
runs a Markov chain at each of the  $n+1$ temperature levels targeting the
joint distribution given by $\prod_{i=0}^{n} [\pi(\theta_i)]^{\beta_i}$. Swap
moves between chains at adjacent temperature levels are accepted according to a
ratio that no longer depends on the marginal normalisation constants. Indeed, this 
power tempering approach has been successfully employed in a number of settings and is 
widely used e.g.\  \cite{Neal1996}, \cite{earl2005parallel}, \cite{Xie2010},  \cite{Mohamed2012} 
and  \cite{Carter2013}.

In both approaches, there is a ``Goldilocks'' principle to setting up the
inverse temperature schedule. Spacings between temperature levels that are 
``too large" result in swap moves that are rarely accepted, hence delaying the transfer of hot
state mixing information to the cold states. On the other
hand, spacings that are too small require a large number of intermediate
temperature levels, again resulting in slow mixing through the
temperature space. This problem becomes even more difficult as the
dimensionality of $\Theta$ increases.

Much of the historical literature suggested that a geometric spacing was
optimal i.e., there exists $c\in (0,1)$ such that  $\beta_{i+1}=c\beta_i$ for
$i=0,1,\ldots,n$. However, in the case of the simulated tempering version (ST), 
\cite{atchade2011towards} considered the problem as
an optimal scaling problem by maximising the (asymptotic in dimension) expected
squared jumping distance in the $B$ space for temperature swap moves.  Under
restrictive assumptions, he showed that the spacings
between consecutive inverse temperature levels should scale with dimension as
$O\left(d^{-1/2}\right)$ to prevent degeneracy of the swap move acceptance
rate. For a practitioner the result gave guidance on optimal setup since it
suggested a corresponding optimal swap move acceptance rate of 0.234 between
consecutive inverse temperature levels, in accordance with
\cite{gelman:gilks:roberts:1996}. Finally, contrary to the historically
recommended geometric schedule, the authors suggested that temperature schedule setup
should be constructed consecutively so as to induce an approximate 0.234 swap
acceptance rate between consecutive levels; which is achieved adaptively in
\cite{miasojedow2013adaptive}. The use of expected squared jumping distance as
the measure of mixing speed was justified in \cite{roberts2014minimising}
where, under the same conditions as in \cite{atchade2011towards}, it was shown
that the temperature component of the ST chain has an associated diffusion
process.

\begin{figure}[h]
\begin{center}
\includegraphics[height=6cm,width=9cm]{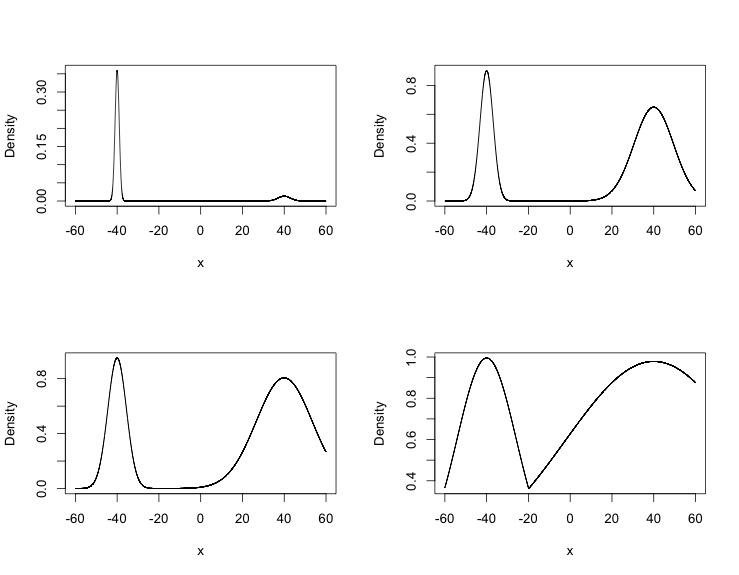}
\caption{Un-normalised tempered target densities of a bimodal Gaussian mixture using inverse temperature levels $\beta=\{ 1,0.1,0.05,0.005 \}$ respectively. At the hot state (bottom right) it is evident that the mode centred on 40 begins to dominate the weight as $\beta$ increases to $\infty$ even though at the cold state it was only attributable for a fraction (0.2) of the total mass.}
\label{Fig:badweight}
\end{center}
\end{figure}

The target of an 0.234 acceptance rate gives good guidance to setting up the ST/PT
algorithms in certain settings, but there is a major warning for practitioners
following this rule for optimal setup.   The assumptions made in
\cite{atchade2011towards} and \cite{roberts2014minimising} ignore the
restrictions of mixing within a temperature level, instead assuming that this
can be done infinitely fast relative to the mixing within the temperature
space. \cite{woodard2009conditions}, \cite{woodard2009sufficient} and \cite{Bhatnagar2016} undertake
a comprehensive analysis of the spectral gap of the ST/PT chains and their
conclusion is rather damning of the ST/PT approaches that use power-tempered
targets. Essentially, in situations where the modes have different structures,
the time required to reach a given level of convergence for the ST/PT
algorithms can grow exponentially in dimension. A major reason for this is that
power-based tempering does not preserve the relative weights/mass between regions at the different 
temperature levels, see Figure~\ref{Fig:badweight}. This issue can scale exponentially in dimension. 
From a practical perspective, in these finite run  high-dimensional non-identical  modal
structure settings the swap acceptance rates can be very misleading, meaning
that they have limited use as a diagnostic for inter-modal mixing quality.

\subsection{Adaptive MCMC}

Improving and calibrating an MCMC algorithm towards a better correspondance with
the intended target is a natural step in making the algorithm more efficient,
provided enough information is available about this target distribution. For
instance, when an MCMC sample associated with this target is available, even
when it has not fully explored the range of the target, it contains some
amount of information, which can then be exploited to construct new MCMC
algorithms. Some of the solutions available in the literature \citep[see,
e.g.][]{liangetal06} proceed by repeating blocks of MCMC iterations and
updating the proposal $K$ after each block, aiming at a particular optimality
goal like a specific acceptance rate like $0.234$ for Metropolis--Hastings
steps \citep{gelman:gilks:roberts:1996}. Most versions of this method update the scale structure of a random walk proposal, based on previous
realisations \citep{robert:casella:2009} or on an entire sample
\citep{douc:guillin:marin:robert:2005}, which turns the method into iterated
importance sampling with Markovian dependence. (It can also be seen as a static
version of particle filtering,
\citealp{doucet:godsill:andrieu:2000,andrieu:doucet:2002,storvik:2002}.) 

\begin{figure}
        \centering
        \includegraphics[width=.6\textwidth,keepaspectratio=true]{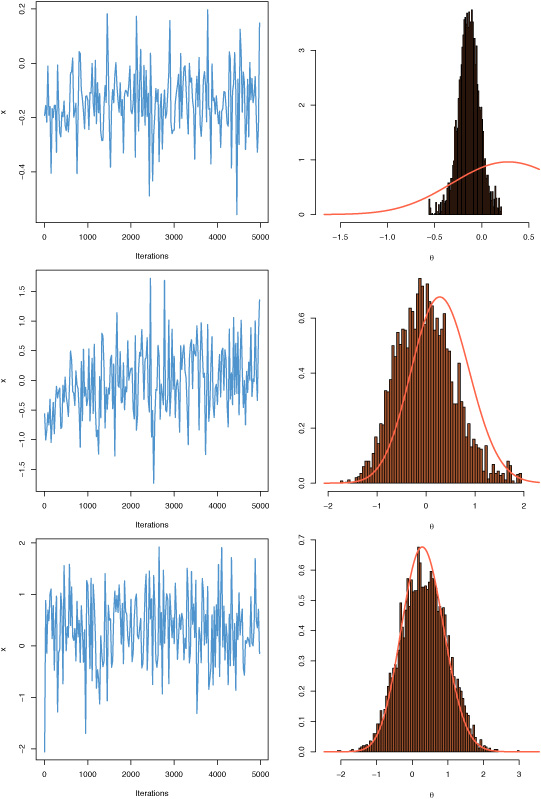}
        \caption{Markov chains produced by an adaptive algorithm where the proposal distribution is a Gaussian distribution with mean and variance computed from the past simulations of the chain. The three rows correspond to different initial distributions. The fit of the histogram of the resulting MCMC sample is poor, even for the most spread-out initial distribution (bottom) \citep[with permission]{robert:casella:2004}}
        \label{figadap}
\end{figure}

Other adaptive resolutions bypass this preliminary and somewhat {\em ad hoc}
construction and aim instead at a permanent updating within the algorithm,
motivated by the idea that a continuous adaptation keeps improving the
correspondance with the target.  In order to preserve the validation of the method
\citep{gelman:gilks:roberts:1996,haario:sacksman:tamminen:1999,roberts:rosenthal:2007,saksman:vihola:2010},
namely that the chain produced by the algorithm converges to the intended
target, specific convergence results need be established, as the ergodic
theorem behind standard MCMC algorithms do not apply. Without due caution (see Figure 
\ref{figadap}), an
adaptive MCMC algorithm may fail to converge due to over-fitting. A drawback of
adaptivity is that the update of the proposal distribution relies {\em too
much} on the earlier simulations and thus reinforces the exclusion of parts of
the space that have not yet been explored.

For the validation of adaptive MCMC methods, stricter constraints must thus be imposed on
the algorithm. One well-described solution \citep{roberts:rosenthal:2006} is
called {\em diminishing adaptation}. Informally, it consists in imposing a distance between
two consecutive proposal kernels to uniformly decrease to zero. In practice, this means
stabilising the changes in the proposal by ridge-like factors as in the early proposal by
\cite{haario:sacksman:tamminen:1999}. A drawback of this resolution is that the
decrease itself must be calibrated and may well fail to bring a significant
improvement over the original proposal.

\subsection{Multiple try MCMC}

A completely different approach to improve the original proposal used in an
MCMC algorithm is to consider a collection of proposals, built on different
rationales and experiments. The {\em multiple try MCMC algorithm}
\citep{liu:liang:wong:2000,bedard:douc:moulines:2012,martino:2018} follows this perspective.
As the name suggests, the starting point of a multiple try MCMC algorithm is to
simultaneously propose $N$ potential moves $\theta^1_t,\ldots,\theta^N_t$ of the Markov chain, instead of a
single value. The proposed values $\theta^i_t$ may be independently generated according to
$N$ different proposal densities $K_i(\cdot|\theta_t)$ that are conditional on the current value of
the Markov chain, $\theta_t$.  One of the $\theta^i_t$'s is selected based on the importance
sampling weights $w_t^i\propto \pi(\theta^i_t)/K_i(\cdot|\theta_t)$. The
selected value is then accepted by a further Metropolis--Hastings step which involves
a ratio of normalisation constants for the importance stage, one corresponding to the selection made
previously and another one created for this purpose. Indeed, besides the added
cost of computing the sum of the importance weights and generating the
different variates, this method faces the non-negligible drawback of requiring
$N-1$ supplementary simulations that are only used for achieving detailed balance
and computing a backward summation of importance weights. This constraint may vanish when
considering a collection of independent Metropolis-Hastings proposals, $q(\theta)$,
but this setting is rarely realistic as it requires
which make life simpler, but are less realistic since 
some amount of prior knowledge or experimentation to build a relevant distribution.

\begin{figure}[H]
        \centering
        \includegraphics[width=.6\textwidth,keepaspectratio=true]{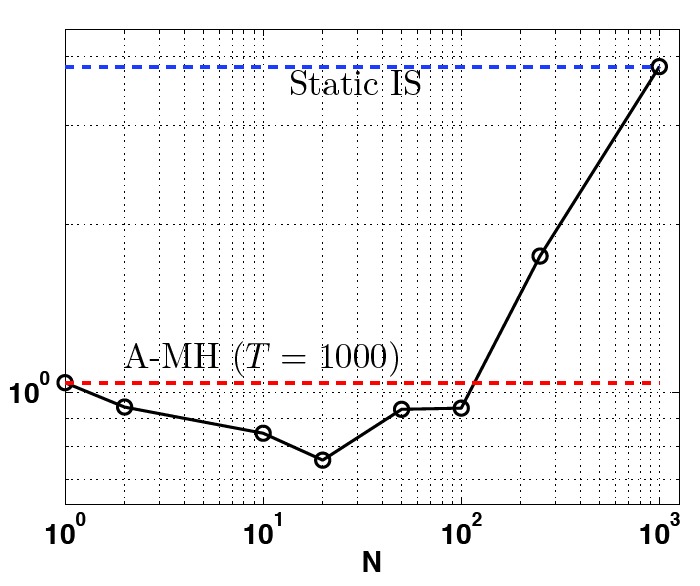}
        \caption{A comparison of an ensemble MCMC approach with a regular adaptive MCMC algorithm (lower line) and a static importance sampling approach, in terms of mean square error (MSE), for a fixed total number of likelihood evaluations, where $N$ denotes the size of the ensemble ({\em Source:} \citealp{martino:2018}, with permission).}
        \label{figensemble}
\end{figure}

An alternative found in the literature is {\em ensemble Monte Carlo} \citep{iba:2000,
cappe:douc:guillin:marin:robert:2007,neal:2011e,martino:2018}, illustrated in Figure \ref{figensemble}
which produces a whole sample at each iteration, with target the product
of the initial targets, in closer proximity with particle methods
\citep{cappe:guillin:marin:robert:2003,mengersen:robert:2003}.

Yet another implementation of this principle is called
{\em delayed rejection} \citep{tierney:mira:1998,mira:2001,miraetal01}, where proposals are instead
considered sequentially, once the previous proposed value has been rejected.
to speeding up MCMC by considering several possibilities, if sequentially.
A computational difficulty with this approach is that the associated acceptance
probabilities get increasingly complex as the number of delays grows, which may annihilate its
appeal relative to simultaneous multiple tries. A further difficulty is to devise the sequence of proposals in a diverse enough manner.

\subsection{Multiple proposals and parameterisations}

A rather basic approach to comparing proposals of MCMC algorithms is to run
several in parallel and to check whether these parallel chains can be exchanged
by coupling. Chains with divergent behaviour will not couple as often as chains
exploring the same area. While creating multiple MCMC algorithms may seem a
major challenge, automated and semi-automated schemes can be replicated as much
as desired by changing the parameterisation of the target. Each change
introduces a different Jacobian in the expression of the density, which means
different efficiencies in the exploration of the target.

\section{Accelerating MCMC by reducing the variance}\label{sec:reduce}

Since the main goal of MCMC is to produce approximations for quantities of interest of the form
$$
\mathfrak{I}_h = \int_\Theta h(\theta) \pi(\theta) \text{d}\theta,
$$
an alternative (and cumulative) way of accelerating these algorithms is to
improve the quality of the approximation derived from an MCMC output. That is,
given an MCMC sequence $\theta_1,\ldots,\theta_T$, converging to
$\pi(\cdot)$, one can go beyond resorting to the basic Monte Carlo
approximation
\begin{equation}\label{eq:bazmo}
\hat{\mathfrak{I}}_h^T = \nicefrac{1}{T}\sum_{t=1}^T h(\theta_t)
\end{equation}
towards reducing the variance (if not the speed of convergence) of
$\hat{\mathfrak{I}}_h^T$ to ${\mathfrak{I}}_h$.

A common remark when considering Monte Carlo approximations of $\mathfrak{I}_h$
is that the representation of the integral as an expectation is not unique
\citep[e.g.][]{robert:casella:2004}. This leads to the technique of importance
sampling where alternative distributions are used in replacement of
$\pi(\theta)$, possibly in an adaptive manner
\citep{douc:guillin:marin:robert:2007a}, or sequentially as in particle filters
\citep{delmoral:doucet:jasra:2006,andrieu:doucet:holenstein:2010}. Within the
framework of this essay, the outcome of a given MCMC sampler can also be
exploited in several ways that lead to an improvement of the approximation of
$\mathfrak{I}_h$.

\subsection{Rao--Blackwellisation and other averaging techniques}

The name `Rao--Blackwellisation' was coined by 
\cite{gelfand:smith:1990} in their foundational Gibbs sampling paper and it
has since then become a standard way of reducing the variance of integral
approximations. While it essentially proceeds from the basic probability
identity
$$\mathbb{E}^\pi[h(\theta)]=\mathbb{E}^{\pi_1}[\mathbb{E}^{\pi_2}\{h(\theta)|\xi\}],$$
when $\pi$ can be expressed as the following marginal density
$$\pi(\theta)=\int_{\Xi} \pi_1(\xi)\pi_2(\theta|\xi)\text{d}\xi\,,$$
and while sufficiency does not have a clear equivalence for Monte Carlo
approximation, the name stems from the Rao--Blackwell theorem
\citep{lehmann:casella:1998} that improves upon a given estimator by
conditioning upon a sufficient statistics. In a Monte Carlo setting, this means
that \eqref{eq:bazmo} can be improved by a partly integrated version
\begin{equation}\label{eq:razmo}
\tilde{\mathfrak{I}}_h^T = \nicefrac{1}{T}\sum_{t=1}^T \mathbb{E}^{\pi_2}[h(\theta)|\xi^t]
\end{equation}
assuming that a second and connected sequence of simulations $(\xi_t)$ is available and that the conditional expectation is easily constructed. For instance, Gibbs sampling \citep{gelfand:smith:1990} is often open to this Rao--Blackwell decomposition as it relies on successive simulations from several conditional distributions, possibly including auxiliary variates and nuisance parameters. In particular, a generic form of Gibbs sampling called the slice sampler \citep{robert:casella:2004} produces one or several uniform variates at each iteration. 

However, a more universal type of Rao--Blackwellisation is available
\citep{casella:robert96} for all MCMC methods involving rejection, first and
foremost, Metropolis--Hastings algorithms. Indeed, first, the distribution of
the rejected variables can be derived or approximated, which leads to an
importance correction of the original estimator. Furthermore, the accept-reject
step depends on a uniform variate, but this uniform variate can be integrated
out. Namely, given a sample produced by a Metropolis--Hastings algorithm
$\theta^{(1)},\ldots,\theta^{(T)}$, one can exploit both underlying samples,
the proposed values $\vartheta_1,\ldots,\vartheta_T$, and the uniform
$u_1,\ldots,u_T$, so that the ergodic mean can be rewritten as
$$
\hat{\mathfrak{I}}_h^T = \nicefrac{1}{T} \sum_{t=1}^T h(\theta^{(t)}) 
= \nicefrac{1}{T} \sum_{t=1}^T h(\vartheta_t) \sum_{i=t}^T \mathbb{I}_{\theta^{(i)}=\vartheta_t}\,.
$$
The conditional expectation
\begin{align*}
\tilde{\mathfrak{I}}_h^T &= \nicefrac{1}{T} \sum_{t=1}^T h(\vartheta_t)
\mathbb{E}\left[\sum_{i=t}^T \mathbb{I}_{\theta^{(i)}=\vartheta_t}\bigg|\vartheta_1,\ldots,\vartheta_T\right]\\
&= \nicefrac{1}{T} \sum_{t=1}^T h(\vartheta_t) \left\{
\sum_{i=t}^T \mathbb{P}(\theta^{(i)}=\vartheta_t|\vartheta_1,\ldots,\vartheta_T) \right\}
\end{align*}
then enjoys a smaller variance. See also \cite{tjemeland:2004} and \cite{douc:robert:2010} for connected improvements based on multiple tries. An even more rudimentary (and cheaper) version can be considered by integrating out the decision step at each Metropolis--Hastings iteration: if $\theta_t$ is the current value of the Markov chain and $\vartheta_t$ the proposed value, to be accepted (as $\theta_{t+1}$) with probability $\alpha_t$, the version
$$
\nicefrac{1}{T} \sum_{t=1}^T \left\{ \alpha_t h(\vartheta_t) +(1-\alpha_t) h(\theta_t) \right\}
$$
should most often\footnote{The improvement is not universal, due to the correlation between the terms of the sum induced by the Markovian nature of the sequence $\{\theta_t\}_{t=1}^T$.} bring an improvement over the basic estimate \citep{liu:won:kon95,robert:casella:2004}.


\section{Conclusion}

Accelerating MCMC algorithms may sound like a new Achille versus tortoise paradox in that there are aways methods to speed up a given algorithm. The stopping rule of this infinite regress is however that the added pain in achieving this acceleration may overcome the added gain at some point. While we have only and mostly superficially covered some of the possible directions in this survey, we thus encourage most warmly readers to keep an awareness for the potential brought by a wide array of almost cost-free accelerating solutions as well as to keep trying devising more fine-tuned improvements in every new MCMC implementation. For instance, for at least one of us, Rao-Blackwellisation is always considered at this stage. Keeping at least one such bag of tricks at one's disposal is thus strongly advised.

\section*{Acknowledgements}

Christian P.~Robert is grateful to Gareth Roberts, Mike Betancourt, and Julien Stoehr for helpful discussions. He is currently supported by an Institut Universitaire de France 2016--2021 senior grant. Changye Wu is currently a Ph.D. candidate at Universit\'e Paris-Dauphine and supported by a grant of the Chinese Government (CSC). V\'ictor Elvira  acknowledges support from the \emph{Agence Nationale de la Recherche} of France under PISCES project (ANR-17-CE40-0031-01), the Fulbright program, and the Marie Curie Fellowship (FP7/2007-2013) under REA grant agreement n. PCOFUND-GA-2013-609102, through the PRESTIGE program. The authors are quite grateful to a reviewer for his or her detailed coverage of an earlier version of the paper, which contributed to significant improvements in the presentation and coverage of the topic. All remaining errors and ommissions are solely the responsability of the authors.


\input arXclmc.bbl

\end{document}

%% file: arXclmc.bbl
\hyphenation{Post-Script Sprin-ger}